\documentclass[aps,prc,twocolumn,superscriptaddress,showpacs,twoside,floatfix,longbibliography]{revtex4-2}

\usepackage{amssymb,epsfig}
\usepackage{mathtools}
\usepackage[normalem]{ulem}
\usepackage{graphicx}
\usepackage{dcolumn}
\usepackage{bm}
\usepackage{multirow}
\usepackage{booktabs}
\usepackage{orcidlink}
\usepackage{hyperref}
\hypersetup{breaklinks=true,colorlinks=true,linkcolor=blue,citecolor=blue,filecolor=magenta,urlcolor=blue}
\usepackage{xcolor}

\begin{document}

\title{Ground State Decay of the Three-Proton Emitter $^{17}$Na \\ Reveals Isospin Symmetry Breaking}

\author{X.-D.~Xu}
\email{Contact author: xiaodong.xu@impcas.ac.cn}
\affiliation{Institute of Modern Physics, Chinese Academy of Science, Lanzhou 730000, China}
\affiliation{School of Nuclear Science and Technology, University of Chinese Academy of Sciences, Beijing 100049, China}

\author{I.~Mukha}
\email{Contact author: I.Mukha@gsi.de}
\affiliation{GSI Helmholtzzentrum f\"{u}r Schwerionenforschung GmbH, 64291 Darmstadt, Germany}

\author{Z. C.~Xu}
\affiliation{Key Laboratory of Nuclear Physics and Ion-beam Application (MOE), Institute of Modern Physics, Fudan University, Shanghai 200433, China}
\affiliation{Shanghai Research Center for Theoretical Nuclear Physics, NSFC and Fudan University, Shanghai 200438, China}

\author{S. M.~Wang}
\email{Contact author: wangsimin@fudan.edu.cn}
\affiliation{Key Laboratory of Nuclear Physics and Ion-beam Application (MOE), Institute of Modern Physics, Fudan University, Shanghai 200433, China}
\affiliation{Shanghai Research Center for Theoretical Nuclear Physics, NSFC and Fudan University, Shanghai 200438, China}

\author{K. Y.~Zhang}
\email{Contact author: zhangky@caep.cn}
\affiliation{National Key Laboratory of Neutron Science and Technology, Institute of Nuclear Physics and Chemistry, China Academy of Engineering Physics, Mianyang, Sichuan 621900, China}

\author{L.~Acosta}
\affiliation{Instituto de Estructura de la Materia, Consejo Superior de Investigaciones Científicas, 28006, Madrid, Spain}

\author{E.~Casarejos}
\affiliation{CINTECX, Universidade de Vigo, DSN, Dpt. Mech. Engineering, E-36310 Vigo, Spain}

\author{D.~Cortina-Gil}
\affiliation{Instituto de Física Corpuscular, CSIC -Universidad de Valencia, 46980, Paterna, Valencia, Spain}

\author{J.~M.~Espino}
\affiliation{Department of Atomic, Molecular and Nuclear Physics, University of Seville, 41012 Seville, Spain}

\author{A.~Fomichev}
\affiliation{Flerov Laboratory of Nuclear Reactions, JINR, 141980 Dubna, Russia}

\author{H.~Geissel}
\altaffiliation{Deceased.}
\affiliation{GSI Helmholtzzentrum f\"{u}r Schwerionenforschung GmbH, 64291 Darmstadt, Germany}
\affiliation{II.Physikalisches Institut, Justus-Liebig-Universit\"at, 35392 Gie{\ss}en, Germany}

\author{J.~G\'{o}mez-Camacho}
\affiliation{Department of Atomic, Molecular and Nuclear Physics, University of Seville, 41012 Seville, Spain}

\author{L.V.~Grigorenko}
\affiliation{Flerov Laboratory of Nuclear Reactions, JINR, 141980 Dubna, Russia}
\affiliation{National Research Nuclear University ``MEPhI'', 115409 Moscow, Russia}
\affiliation{National Research Centre ``Kurchatov Institute'', Kurchatov square 1, 123182 Moscow, Russia}

\author{O.~Kiselev}
\affiliation{GSI Helmholtzzentrum f\"{u}r Schwerionenforschung GmbH, 64291 Darmstadt, Germany}

\author{A.A.~Korsheninnikov}
\affiliation{National Research Centre ``Kurchatov Institute'', Kurchatov square 1, 123182 Moscow, Russia}

\author{N.~Kurz}
\affiliation{GSI Helmholtzzentrum f\"{u}r Schwerionenforschung GmbH, 64291 Darmstadt, Germany}

\author{Yu.A.~Litvinov}
\affiliation{GSI Helmholtzzentrum f\"{u}r Schwerionenforschung GmbH, 64291 Darmstadt, Germany}
\affiliation{Institut f\"{u}r Kernphysik, Universit\"{a}t zu K\"{o}ln, D-50937 K\"{o}ln, Germany}

\author{I.~Martel}
\affiliation{University of Huelva, 21007 Huelva, Spain}

\author{C.~Nociforo}
\affiliation{GSI Helmholtzzentrum f\"{u}r Schwerionenforschung GmbH, 64291 Darmstadt, Germany}

\author{M.~Pf\"{u}tzner}
\affiliation{Faculty of Physics, University of Warsaw, 02-093 Warszawa, Poland}

\author{C.~Rodr\'{i}guez-Tajes}
\affiliation{Universidade de Santiago de Compostela, 15782 Santiago de Compostela, Spain}

\author{C.~Scheidenberger}
\affiliation{GSI Helmholtzzentrum f\"{u}r Schwerionenforschung GmbH, 64291 Darmstadt, Germany}
\affiliation{II.Physikalisches Institut, Justus-Liebig-Universit\"at, 35392 Gie{\ss}en, Germany}
\affiliation{Helmholtz Research Academy Hesse for FAIR (HFHF), GSI Helmholtz Center for Heavy Ion Research, Campus Gießen, 35392 Gießen, Germany}

\author{M.~Stanoiu}
\affiliation{IFIN-HH, Post Office Box MG-6, Bucharest, Romania}

\author{K.~S\"{u}mmerer}
\affiliation{GSI Helmholtzzentrum f\"{u}r Schwerionenforschung GmbH, 64291 Darmstadt, Germany}

\author{H.~Weick}
\affiliation{GSI Helmholtzzentrum f\"{u}r Schwerionenforschung GmbH, 64291 Darmstadt, Germany}

\author{P.J.~Woods}
\affiliation{University of Edinburgh, EH1 1HT Edinburgh, United Kingdom}

\author{M.V.~Zhukov}
\affiliation{Department of Physics, Chalmers University of Technology, S-41296 G\"oteborg, Sweden}

\date{\today}

\begin{abstract}

The spectrum of the exotic three-proton (3\textit{p}) emitter $^{17}$Na has been studied by detecting all in-flight decay products. Derived from the measured angular correlations $^{14}$O+\textit{p}+\textit{p}+\textit{p}, a resonant peak has been discovered at the 3\textit{p}-decay energy of 2.24($^{+0.17}_{-0.25}$)~MeV, which likely corresponds to the $^{17}$Na ground state. This decay energy value is significantly smaller than the previous experimental upper limit.~Our measured $^{14}$O\text{--}\textit{p} correlations stemming from the ground state decay have been quantitatively described by a sequential 1\textit{p}\text{--}2\textit{p} emission from a $^{17}$Na resonance via the intermediate $^{16}$Ne ground state, which allowed to derive the upper limit of $^{17}$Na ground-state width of 0.6~MeV. A dramatic systematic decrease in the mirror energy differences of mirror nuclei pairs has been observed at almost all 3\textit{p} emitters with known proton separation energy (such as $^{31}$K, $^{20}$Al, and $^{17}$Na), in sharp contrast to the behavior in less exotic nuclei.~Such a lowering effect indicates a general trend in evolution of nuclear structure for light to medium mass nuclei beyond the proton drip line, which is often associated with strong isospin symmetry breaking.

\end{abstract}

\maketitle

Isospin symmetry introduced under the assumption of charge symmetry and charge independence of the nuclear force \cite{Wigner:1937} is a fundamental concept in nuclear physics which declares that the spins and parities of the ground states (g.s.)~of mirror nuclei (namely, nuclei with reversed numbers of neutrons and protons) should be the same. Isospin symmetry is almost preserved in all bound nuclear systems.~However, it may be broken due to the Coulomb interaction among protons, the proton and neutron mass difference, and the presence of the charge dependence of the nuclear force.~Exceptions from the isospin symmetry are intriguing, particularly in proton-unbound nuclei.~For example, pronounced isospin symmetry breaking has been revealed between \textit{p}-unbound nucleus $^{16}$F and its mirror partner $^{16}$N~\cite{Stefan:2014}.~Although the structure of 1\textit{p} and 2\textit{p} emitters was addressed in a number of studies (see the recent review in Ref.~\cite{Pfutzner:2023}), the understanding of more exotic 3\textit{p}-unbound nuclei is very limited.~To date, only a few light isotopes have been identified experimentally as 3\textit{p} emitters, including $^{7}$B~\cite{Charity:2011}, $^{13}$F~\cite{Charity:2021}, $^{17}$Na~\cite{Brown:2017}, $^{20}$Al~\cite{Xu2025PRL}, and $^{31}$K~\cite{Kostyleva:2019}.~The measured 3\textit{p} decay patterns of all these nuclei demonstrated a sequential 1\textit{p}\text{--}2\textit{p} decay mechanism.~The g.s.~of $^{17}$Na has not been firmly established. The reported energy spectrum for $^{17}$Na led to an upper limit of 4.85(6)~MeV for its g.s.\ decay energy~\cite{Brown:2017}. 

In this Letter, we present the previously unidentified g.s.\ of 3\textit{p}-emitter $^{17}$Na using data obtained from the experiment with a $^{17}$Ne secondary beam. Based on the g.s.\ decay energy of $^{17}$Na and other 3\textit{p} emitters, we lay out the systematics of the mirror energy difference MED (defined as MED=$S_n-S_p$, where $S_n$ and $S_p$ are the neutron separation energy of a \textit{n}-rich nucleus and the proton separation energy of its \textit{p}-rich mirror partner, respectively). For the first time, we demonstrate the general trend of a dramatic decrease in the MED values of 3\textit{p} emitters, in comparison with \textit{p}-bound nuclei, and theoretically address the underlying physics reasons.

The experiment was described in detail in Refs.~\cite{Mukha:2010,Mukha:2012}. The $^{17}$Ne secondary beam was produced by the fragmentation of a primary 591 \textit{A}MeV $^{24}$Mg beam at the SIS-FRS facility at GSI, Germany. This $^{17}$Ne beam, with an energy of 410 \textit{A}MeV and an intensity of 800 ions $\text{s}^{-1}$, was then transported to bombard a 2 g/cm$^{2}$ $^9$Be secondary target, see details in Refs.~\cite{Mukha:2007,Mukha:2010}. The $^{17}$Na ions were produced via a charge-exchange reaction. The decay products of the unbound $^{17}$Na nuclei were tracked by a double-sided silicon microstrip detector (DSSD) array, which consisted of four large-area DSSDs~\cite{Stanoiu:2008}, placed just downstream of the secondary target. The DSSDs were employed to measure hit coordinates of the protons and the heavy-ion (HI) decay products, resulting from the in-flight decays of the proton precursors. The high-precision position measurements by DSSDs allowed for reconstruction of all fragment trajectories, enabling us to derive a decay vertex together with angular HI-\textit{p} and HI-\textit{p}-\textit{p} correlations. Based on the trajectories of $^{14}$O+$p_1$+$p_2$ measured in coincidence, the relative angles between each proton and $^{14}$O (i.e., $\theta_{p-^{14}\rm{O}}$) were derived. Several states including the $^{16}$Ne g.s.\ were observed, and spectroscopic information on these states was obtained~\cite{Mukha:2010}.

\begin{figure}[!htbp]
\begin{center}
\includegraphics[width=0.45\textwidth]{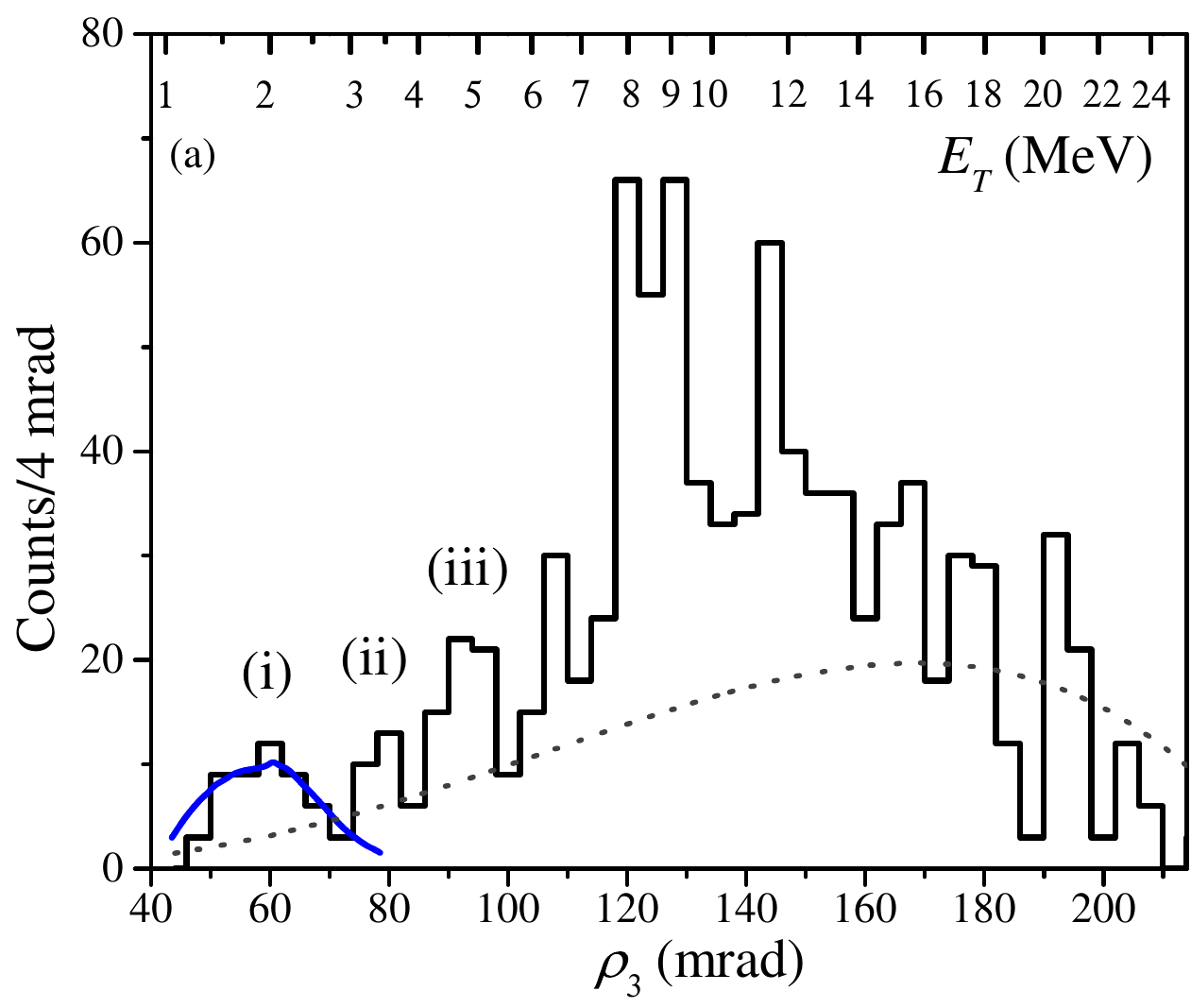}
\includegraphics[width=0.45\textwidth]{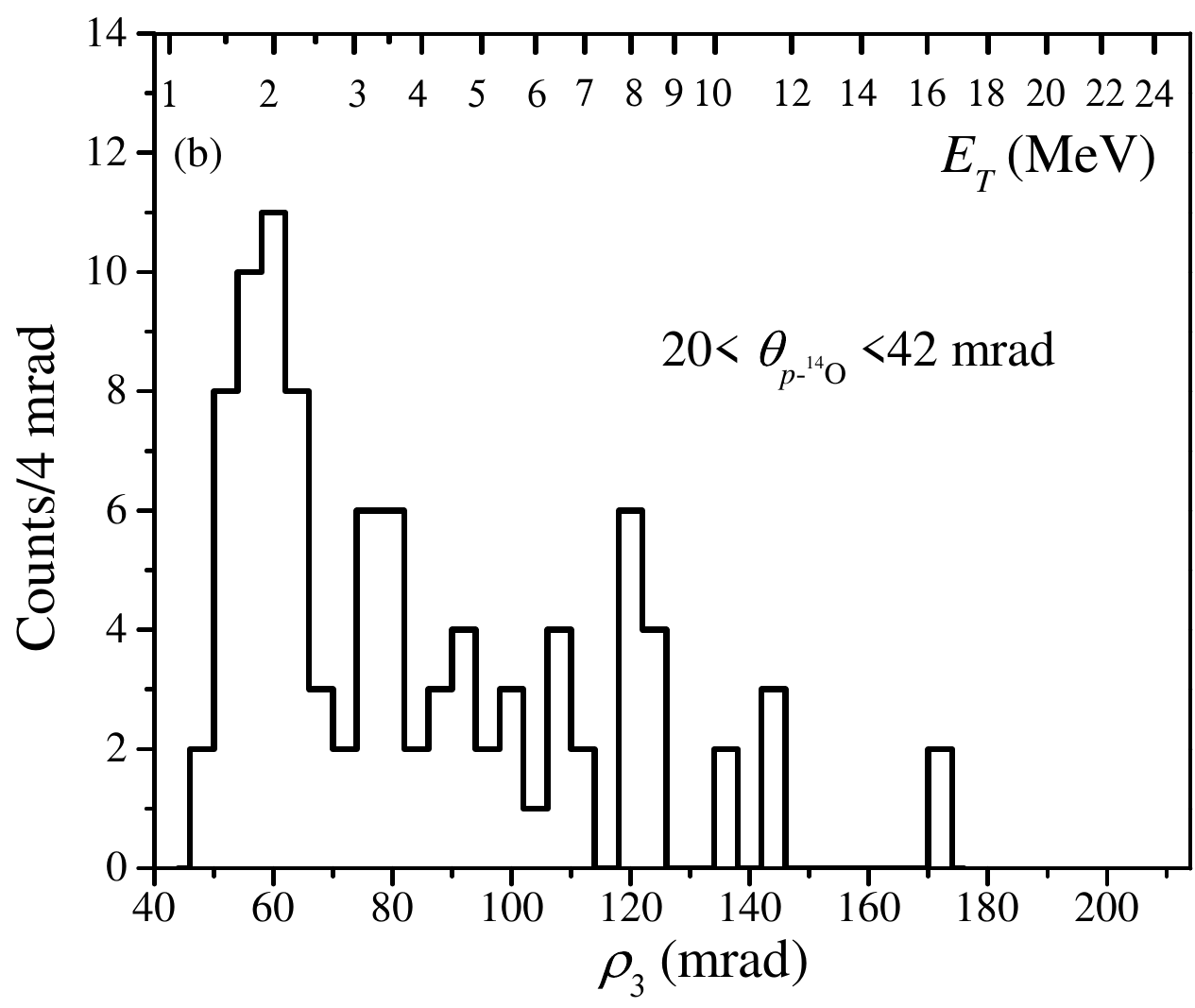}
\end{center}
\caption{(a) Three-proton angular correlations $\rho_3$ derived from the measured trajectories of the decay products of $^{17}$Na, $^{14}$O+3\textit{p} (histogram), which reflect the total 3\textit{p}-decay energy $E_T$ of the $^{17}$Na states shown in the upper axis. The illustrative simulation of the 3\textit{p} decays of $^{17}$Na with the assumed $E_T$ of 2~MeV is shown by the solid curve in region (i). The dotted curve shows a four-body phase volume simulation for a direct reaction with an exit channel $^{14}$O+3\textit{p} in the absence of any resonance in $^{17}$Na. (b) Similar distribution as in (a) but gated by small angles $20<\theta_{p-^{14}\rm{O}}<42$ mrad, which are typical for the $^{16}$Ne g.s.\ decay.}
\label{fig:17Na-rho3}
\end{figure}

The $^{17}$Na 3\textit{p}-decay energy was derived from the trajectories and angular correlations of all decay products $^{14}$O+$p_1$+$p_2$+$p_3$, which were measured in four-fold coincidence. The procedures similar to those applied in the previous $^{16}$Ne studies~\cite{Mukha:2007,Mukha:2010} were used to analyze the $^{14}$O+3\textit{p} correlations and derive the $^{17}$Na spectrum. In particular, the same detector calibration coefficients were employed. The data-analysis procedure utilized in studies of 3\textit{p} decays of $^{31}$K~\cite{Kostyleva:2019} and $^{20}$Al~\cite{Xu2025PRL} was applied to $^{17}$Na. The measured trajectories of coincident $^{14}$O+3\textit{p} events allowed us to derive angles between each proton and the $^{14}$O ions. Then a kinematic variable $\rho_3$ was determined as follows: $\rho_3=\sqrt{\theta^2_{p_1-^{14}\rm{O}}+\theta^2_{p_2-^{14}\rm{O}}+\theta^2_{p_3-^{14}\rm{O}}}$. The emitted protons share the 3\textit{p}-decay energy, making the $\rho_3$ distribution very useful for illustration of the states populated in the 3\textit{p}-decay precursor, as demonstrated in the spectroscopy of $^{31}$K~\cite{Kostyleva:2019} and $^{20}$Al~\cite{Xu2025PRL}.

The $\rho_3$ distribution derived from the measured $^{14}$O+3\textit{p} correlations following $^{17}$Na decays is shown in Fig.~\ref{fig:17Na-rho3}(a). The corresponding total 3\textit{p} decay energies $E_T$ can be estimated from the upper axis.~To evaluate the possible contribution of the nonresonant branch into the measured correlations, we performed the four-body phase volume simulation for a direct reaction with an exit channel $^{14}$O+3\textit{p} in the absence of any resonance in $^{17}$Na. The four-body phase volume is proportional to an $E_T^{7/2}$ factor~\cite{Muzalevskii2025PRC} multiplied to the detection efficiency of $^{14}$O+3\textit{p} events. It is normalized to the measured intensity at small and large $\rho_3$ values, $\le$50 and $\ge$190 mrad, respectively. Due to this normalization, the phase-volume component represents an upper-limit estimate of possible nonresonant reactions. The low-energy part of the spectrum shown in Fig.~\ref{fig:17Na-rho3}(a) significantly exceeds the estimated nonresonant contribution, thus low energy $^{17}$Na resonance contributions are required to describe these measured angular correlations.~In this Letter, we focus on the lowest states, as suggested by peaks (i), (ii), and (iii) in Fig.~\ref{fig:17Na-rho3}(a), with special attention to the $^{17}$Na g.s.~There are also a few prominent peaks in the $\rho_3$ spectrum above 100 mrad, e.g., located at $E_T$ of $\sim$10 and $\sim$14~MeV, which correspond to highly excited states in $^{17}$Na. 

The measured upper limit of the $^{17}$Na g.s.\ decay energy [4.85(6)~MeV]~\cite{Brown:2017} reduces the inspected angular correlations to the range below 80 mrad [see Fig.~\ref{fig:17Na-rho3}(a)]. The unbound $^{17}$Na g.s.\ has a negative $S_p$ value. Hence it should be open to a sequential 1\textit{p}-2\textit{p} decay mechanism via the intermediate g.s.\ of $^{16}$Ne. The decay pattern of the $^{16}$Ne g.s.\ measured in the same experiment shows the $\theta_{p-^{14}\rm{O}}$ correlations ranging from 20 to 42 mrad, see Ref.~\cite{Mukha:2012}. Thus the measured $\rho_3$ correlations may be exclusively inspected by implementing a selective $\theta_{p-^{14}\rm{O}}$ gate. We produced the exclusive $\rho_3$ distribution by applying the gate in the $\theta_{p-^{14}\rm{O}}$ range of 20\text{--}42 mrad, which is typical for the $^{16}$Ne g.s.\ decay. Consequently, the low-lying states in $^{17}$Na decaying via the $^{16}$Ne g.s.\ should remain relatively unaltered, while higher-energy states which decay in larger $\theta_{p-^{14}\rm{O}}$ ranges should be suppressed.~Figure~\ref{fig:17Na-rho3}(b) shows the gated $\rho_3$ distribution with the prominent peak at $E_T$ of $\sim$2~MeV, which suggests the lowest state in $^{17}$Na decaying via the intermediate $^{16}$Ne g.s.\ For illustration purposes, a possible $^{17}$Na state assumed at $E_T$=2~MeV is shown by the solid curve in Fig.~\ref{fig:17Na-rho3}(a). The corresponding contribution is obtained by the GEANT simulations of the detector response to the decay of interest and the data analysis procedure applied to the $\rho_3$ angular correlations, see details in Ref.~\cite{Mukha:2010}. The peak region of simulations is labeled as (i).

\begin{figure}[!htbp]
\begin{center}
\includegraphics[width=0.45\textwidth]{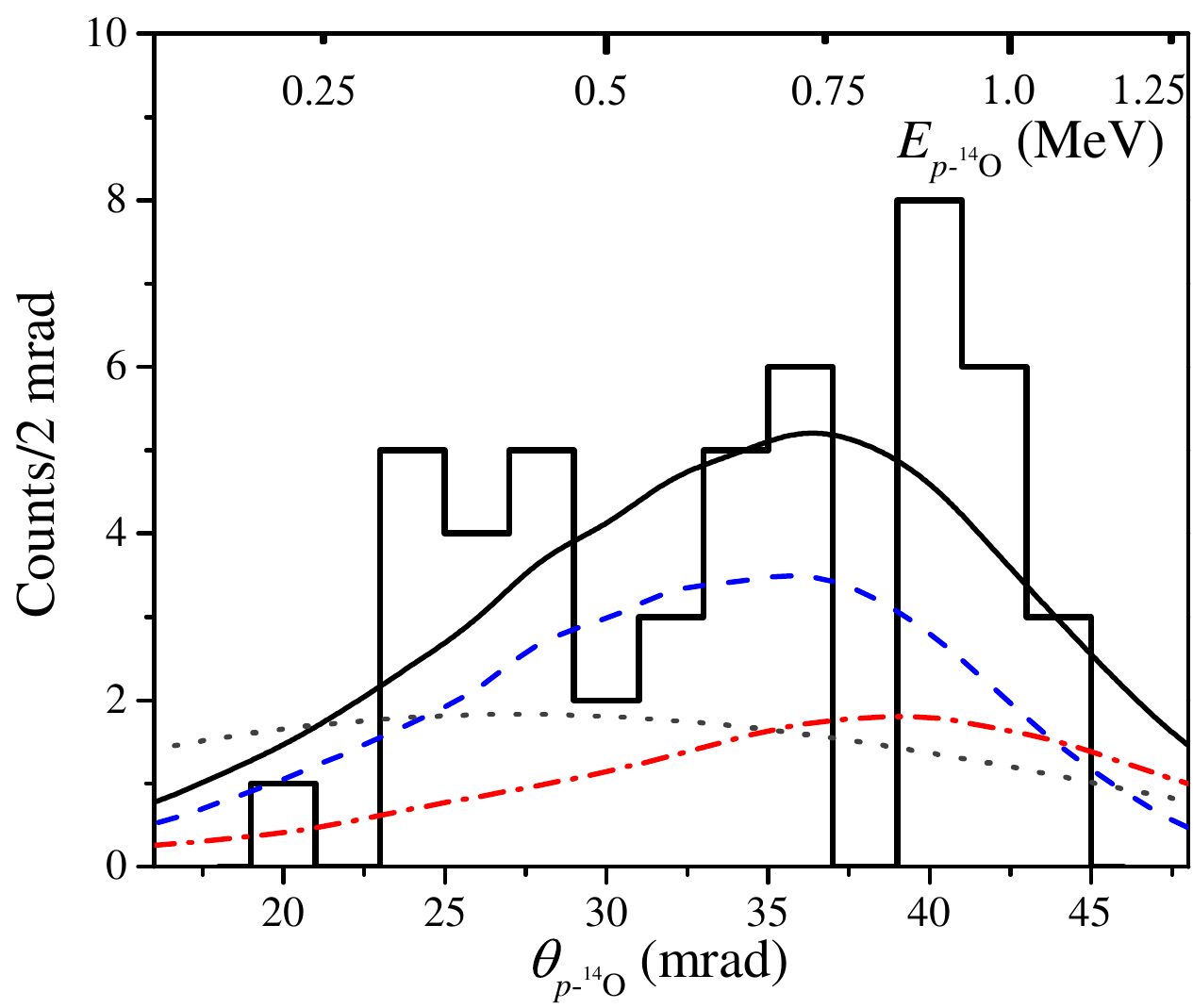}
\end{center}
\caption{Angular $\theta_{p-^{14}\rm{O}}$ correlations (histogram) derived from the measured $^{14}$O+3\textit{p} coincidences using the selection gate (i) shown in Fig.~\ref{fig:17Na-rho3}(a) in the $45<\rho_3<70$ mrad range. The corresponding $1p$-decay energies $E_{p-^{14}\rm{O}}$ are given by the upper axis. The simulated contribution from an initial 1\textit{p} decay of $^{17}$Na into the $^{16}$Ne g.s.\ with the $Q_{1p}$ of 0.84~MeV is shown by the dashed-dotted curve. The contribution of a subsequent 2\textit{p} decay of $^{16}$Ne g.s.\ with the known decay energy of 1.40~MeV~\cite{Mukha:2012} is shown by the dashed curve. The solid curve is their sum corresponding to $E_T$=2.24~MeV. The dotted curve shows a four-body phase volume simulation which represents an upper-limit contribution of nonresonant reactions.}
\label{fig:theta-p-Ne}
\end{figure}

In order to investigate the observed lowest state (i) in $^{17}$Na, which may correspond to the $^{17}$Na g.s., and to establish its decay scheme quantitatively, the events located in the $\rho_3$ region (i) in Fig.~\ref{fig:17Na-rho3}(a) have been selected, and the resulting angular $\theta_{p-^{14}\rm{O}}$ correlations are displayed in Fig.~\ref{fig:theta-p-Ne}.~The angular correlation pattern shows a broad distribution ranging from 20 to 45 mrad with a maximum at a relatively large angle of $\sim$40~mrad. A comparison between Fig.~\ref{fig:17Na-rho3}(a) and Fig.~\ref{fig:17Na-rho3}(b) reveals that the lowest state (i) of $^{17}$Na decays by emission of a proton into the intermediate $^{16}$Ne g.s.\ The 2\textit{p}-decay energy ($Q_{2p}$) of $^{16}$Ne g.s.\ has been measured to be 1.40(2)~MeV~\cite{Mukha:2010,Wang2021_AME}. Therefore, the $\theta_{p-^{14}\rm{O}}$ correlations corresponding to the decay of $^{17}$Na g.s.\ are expected to consist of two contributions.~The first-emitted proton into the intermediate state in $^{16}$Ne is expected to cause a peak in the observed $\theta_{p-^{14}\rm{O}}$ correlations. The secondary 2\textit{p}-component should have the same shape as the known $\theta_{p-^{14}\rm{O}}$ distribution from the $^{16}$Ne g.s.\ 2\textit{p} decay~\cite{Mukha:2010}, which is centered around $E_{p-^{14}\rm{O}}\simeq$ 0.7~MeV (because two identical protons share the total decay energy of 1.4~MeV).

We have evaluated the angular distribution in Fig.~\ref{fig:theta-p-Ne} by summing the two respective components: 1) the simulations of the detector response to the 1\textit{p}-emission of $^{17}$Na with varied  1\textit{p}-decay energy $Q_{1p}$ (the simulation procedure is described in Refs.~\cite{Mukha:2010,Mukha:2012,Xu2025PRL}); 2) the known detector response to the 2\textit{p} decay of $^{16}$Ne g.s.\ (see Ref.~\cite{Mukha:2010}). As shown in Fig.~\ref{fig:theta-p-Ne}, the smaller-angle region of the $\theta_{p-^{14}\rm{O}}$ distribution is described by the 2\textit{p}-decay of $^{16}$Ne g.s.\ (dashed curve) with the known $Q_{2p}$=1.40(2)~MeV. The larger-angle part can be described by the 1\textit{p}-emission of $^{17}$Na into $^{16}$Ne g.s.\ (dash-dotted curve) with the best-matching value $Q_{1p}$=0.84($^{+0.17}_{-0.25}$)~MeV. Together they yield a total 3\textit{p}-decay energy $E_T$=2.24($^{+0.17}_{-0.25}$)~MeV. The data has been compared with the simulations using the standard Kolmogorov-Smirnov (KS) test, which computes the probability that the simulated spectrum matches the respective experimental pattern~\cite{Eadie:1971}.~According to the KS test, two compared histograms are statistical variations of the same distribution if the KS-test probability is larger than 0.5.~As a result, the sum of the above-mentioned two components (solid curve in Fig.~\ref{fig:theta-p-Ne}) matches the data with the largest  probability of 0.89, and the obtained uncertainties of decay energy correspond to the range where the KS-test probability is above 0.5.~Therefore, we conclude that the $^{17}$Na g.s.\ decays predominantly by sequential 1\textit{p}-2\textit{p} emission via the intermediate g.s.\ of $^{16}$Ne. Regarding the width of the $^{17}$Na g.s., we derive an upper limit value $\Gamma_{g.s.}<0.6$~MeV from the data and corresponding descriptions in Fig.~\ref{fig:theta-p-Ne}, which is mainly due to the experimental resolution.

It is interesting to compare the low-energy spectrum obtained in the present work with the corresponding spectrum presented in Ref.~\cite{Brown:2017}.~In both experiments, $^{17}$Na ions were produced using the charge-exchange reactions with $^{17}$Ne ions.~Both experiments derive $^{17}$Na spectra with similar shapes, though differences in the invariant-mass and angular-correlation distributions as well as detection efficiencies must be taken into account. A peak located at $E_T=4.85(6)$~MeV corresponds to the intense peak reported, see Fig.~1 in Ref.~\cite{Brown:2017}. An upper limit of 4.85(6)~MeV for the decay energy of $^{17}$Na g.s.\ was given. Moreover, as noted in Ref.~\cite{Brown:2017}, the assignment of 4.85-MeV peak to a high-lying excited state cannot be ruled out because there are some events located around $E_T\sim3$~MeV. Comparing with our present results, the suggested $^{17}$Na g.s.\ located at $E_T\sim2.2$~MeV overlaps with the events near $E_T\sim3$~MeV shown in Fig.~1 of Ref.~\cite{Brown:2017}. Moreover, the observed hint of a possible state in $^{17}$Na located at $E_T\sim5.2$~MeV overlaps with the right tail of the 4.85-MeV broad peak, see Fig.~1 in Ref.~\cite{Brown:2017}.~One must note different energy calibrations assigned to the derived spectra of $^{17}$Na in the discussed experiments. This difference is larger between high-lying states in $^{17}$Na.

\begin{figure}[hb]
\begin{center}
\includegraphics[width=0.42\textwidth]{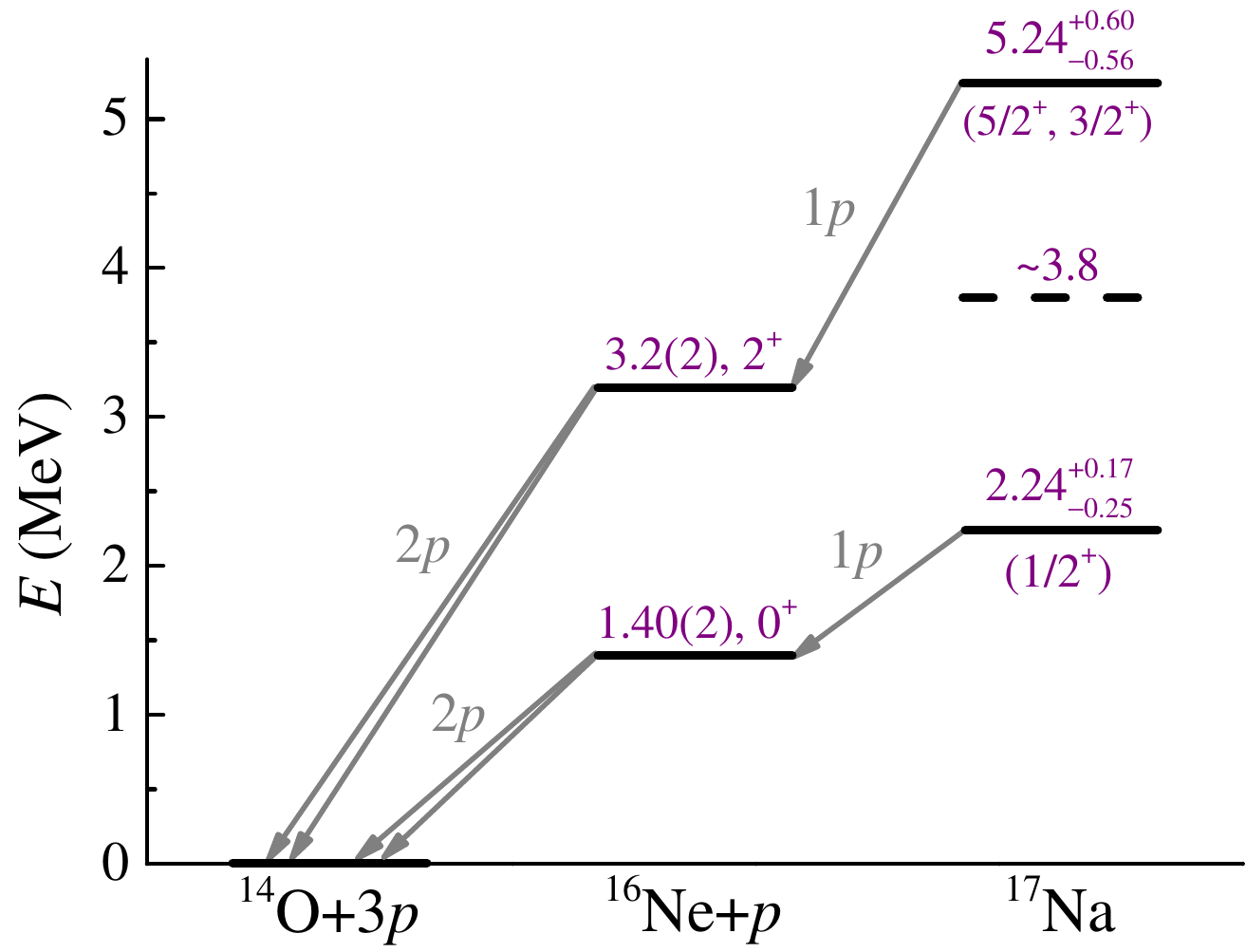}
\end{center}
\caption{Proposed decay scheme of two low-energy states in $^{17}$Na with tentatively assigned decay channels via the known $^{16}$Ne states~\cite{Mukha:2012}, whose energies are given relative to the 3\textit{p} and 2\textit{p} thresholds, respectively. The dashed line at around 3.8~MeV indicates the tentative position of an excited state of $^{17}$Na, i.e., bump (ii) shown in Fig.~\ref{fig:17Na-rho3}(a).}
\label{fig:decay-scheme}
\end{figure}

Besides the g.s., we have evaluated low-energy excited states in $^{17}$Na.~In particular, there are indications of two $^{17}$Na states whose 3\textit{p}-decays produce bumps (ii) and (iii) in the $\rho_3$ spectrum of Fig.~\ref{fig:17Na-rho3}(a), with corresponding $E_T$ of $\simeq$3.8 and $\simeq$5~MeV, respectively. The accumulated statistics in region (ii) is not sufficient, but the 3\textit{p}-decay events in region (iii) allow for a quantitative description. In Appendix A, the angular $\theta_{p-^{14}\rm{O}}$ correlations obtained by imposing the $\rho_3$ gate (iii) in Fig.~\ref{fig:17Na-rho3}(a) are described in the same way as the g.s.-related data in Fig.~\ref{fig:theta-p-Ne}. Consequently, a decay energy of $E_T$=5.24($^{+0.60}_{-0.56}$)~MeV is determined for this excited state in $^{17}$Na, and a spin-parity of (3/2$^+, 5/2^+$) is tentatively assigned.

The derived energies of low-lying states in $^{17}$Na and their decay scheme are displayed in Fig.~\ref{fig:decay-scheme}. Based on the decay energy of the $^{17}$Na g.s.\ and the masses of $^{14}$O+3\textit{p}, the mass excess (M.E.) of $^{17}$Na has been determined to be 32.11(25)~MeV. Such an experimental value is significantly lower than the M.E.\ value of 34.72(6)~MeV tabulated in the AME2020 atomic mass evaluation~\cite{Wang2021_AME}. 
\vspace{-3mm}
\begin{table}[!htbp]
\caption{Theoretical predictions of the proton separation energy $S_p$ and spin-parity $J^\pi$ for the $^{17}$Na g.s.}
\begin{ruledtabular}
  \begin{tabular}{llll}
    & $S_p$ (MeV) & $J^\pi$ & Reference  \\
	\hline
	(a)	& $-$1.03 & ${1/2}^{+}$ & \cite{Amos:2012} \\	
	(b)	& $-$2.40 & ${1/2}^{+}$ & \cite{Timofeyuk2010PRC} \\
	(c)	& $-$2.63 & ${1/2}^{+}$ & \cite{Michel:2021} \\	
    (d)	& $-$2.71 & ${1/2}^{+}$ & \cite{Fortune2010PRC} \\
	(e)	& $-$3.02 & ${1/2}^{+}$ & \cite{Fortune2014PRC} \\
	(f)	& $-3.23\pm0.19$ & \text{---} & \cite{Zong:2022} \\    
	(g)	& $-4.07\pm0.03$\text{*} & \text{---} & \cite{Tian:2013} \\
    (h)	& $-$2.85 & ${5/2}^{+}$ & GSM, this work \\
    (i)	& ($-$2.0, $-$1.0) & (${1/2}^{+}$)\text{**} & GCC, this work \\
    (j)	& $-$1.20 & ${1/2}^{+}$ & DRHBc, this work \\
  \end{tabular}
\end{ruledtabular}
\label{tab:compare}
\text{\hspace{-2mm}*Calculated using the mass predicted for $^{17}$Na in Ref.~\cite{Tian:2013}.}
\text{\hspace{-8mm}**Virtual-like threshold resonance (in terms of GCC).}
\end{table}

The $S_p$ and $J^\pi$ values for the $^{17}$Na g.s.\ were addressed with a number of theoretical models~\cite{Timofeyuk2010PRC,Fortune2010PRC,Amos:2012,Fortune2014PRC,Michel:2021}, and the mass of $^{17}$Na was predicted by several mass systematics, e.g., Refs.~\cite{Tian:2013,Zong:2022}.~These predictions show significant discrepancies, as can be seen in Table~\ref{tab:compare}, which lists theoretical predictions available in literature and three model predictions from the present work.~Among the predictions for $S_p$, those from Ref.~\cite{Amos:2012} and the deformed relativistic Hartree-Bogoliubov theory in continuum (DRHBc)~\cite{Zhou2010PRC(R),Li2012PRC} show approximate agreement with our measured value of $-$0.84(25)~MeV.~Most of the theoretical models predict the spin-parity assignment $J^\pi$=1/2$^+$ for the $^{17}$Na g.s., which is different compared to its mirror nucleus $^{17}$C with the known $J^\pi$=3/2$^+$~\cite{Maddalena2001PRC}. Therefore, one more case of isospin symmetry breaking beyond the proton drip line is suggested.

\begin{figure}[!htbp]
\begin{center}
\includegraphics[width=0.46\textwidth]{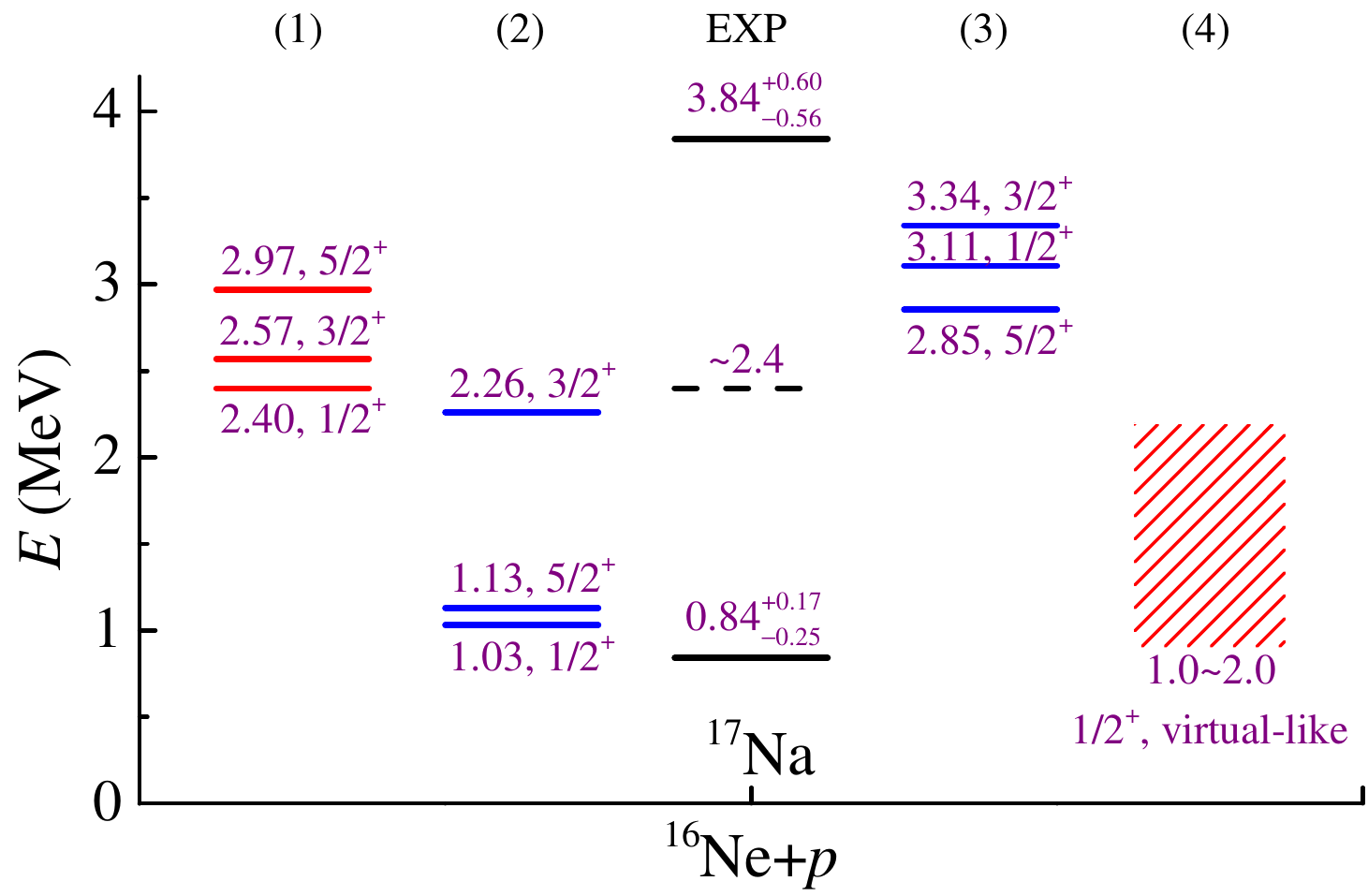}
\end{center}
\caption{The experimental energies of the lowest states in $^{17}$Na observed in the present work compared to various model predictions. Predictions (1) and (2) are taken from Ref.~\cite{Timofeyuk2010PRC} and Ref.~\cite{Amos:2012}, respectively.  Result (3) is calculated using the GSM. Result (4) shows the possible range for the virtual-like threshold resonance estimated by the GCC. The energy values are provided with respect to the 1\textit{p} threshold.}
\label{fig:exp_th_comp}
\end{figure}

We studied low-lying states of $^{17}$Na and its 1\textit{p}-decay daughter $^{16}$Ne with the Gamow shell model (GSM)~\cite{Michel_GSM_book,Michel:2002} and the Gamow coupled-channel (GCC) approach~\cite{Wang:2017,Wang2019}.~A brief description of the GSM and GCC as well as calculation details are in the Appendix B. Figure~\ref{fig:exp_th_comp} compares the GSM and GCC predictions [labeled as (3) and (4), respectively] with the data, along with predictions of two other models~\cite{Timofeyuk2010PRC,Amos:2012}. The cluster  model (2)~\cite{Amos:2012} predicts both 1/2$^+$ g.s.\ and $\sim$2.4-MeV excited state in $^{17}$Na.~The g.s.\ energy is well reproduced within the experimental uncertainty. The model (1)~\cite{Timofeyuk2010PRC} predicts higher energy of the 1/2$^+$ g.s.\ The continuum involved GSM calculations predict a 5/2$^+$ g.s.\ at much higher energy with $S_p$=$-$2.85~MeV. The GCC calculations indicate that the observed $^{17}$Na g.s.\ might correspond to a virtual-like threshold resonance with the decay energy $Q_p\lessapprox2$~MeV, see Appendix B for more details.~It is compelling to find out whether the g.s.\ peak in $^{17}$Na spectrum is due to the virtual-like threshold resonance or due to the individual nuclear structure of $^{17}$Na.~A sensitive parameter here is the resonance width.~While the GCC predicts the width of a virtual-like threshold resonance which is comparable with its energy (i.e. of~1\text{--}2~MeV), a Wigner upper-limit estimate of the width of the $^{17}$Na g.s.\ (with $S_p$=$-$0.84~MeV and $L_p$=0) is $\sim$120~keV. Due to the experimental resolution, the present data allow us only to derive an upper limit value for the $^{17}$Na g.s.\ width $\Gamma_{g.s.}<$ 0.6~MeV. Future measurements with improved statistics and better resolution are required to further elucidate the nature of the $^{17}$Na g.s.

\begin{figure}[t]
\begin{center}
\includegraphics[width=0.45\textwidth]{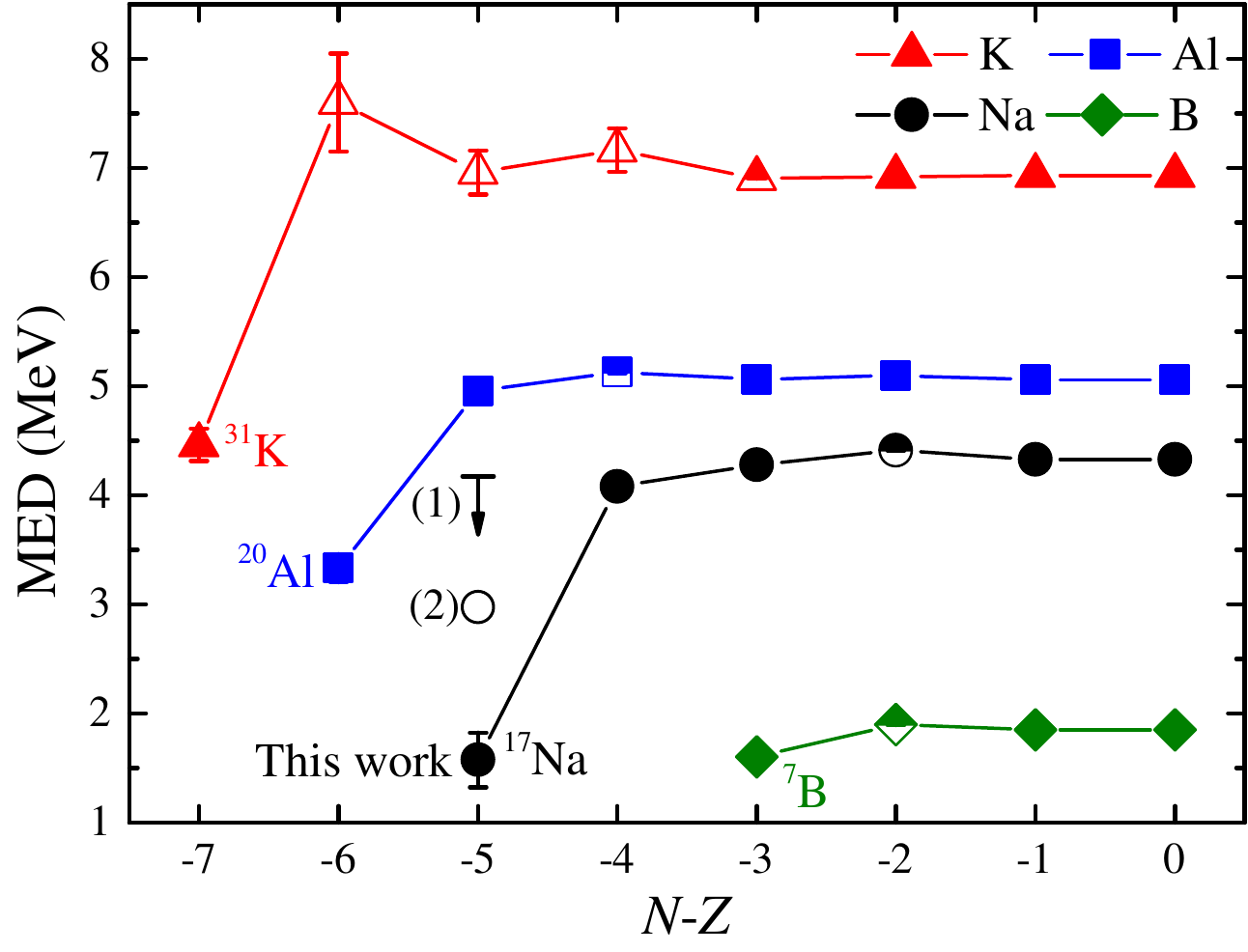}
\end{center}
\caption{Mirror energy differences (MEDs) as a function of $N-Z$ of the \textit{p}-rich partners for B, Na, Al, and K isotopes. All studied 3\textit{p}-emitters ($^{31}$K, $^{20}$Al, $^{17}$Na, and $^{7}$B) deviate from the systematics. The solid symbols are calculated using the measured mass data available in AME2020~\cite{Wang:2021}. The hollow triangles are computed employing the extrapolated mass values for $^{32,33,34}$K from AME2020. The half-filled symbols represent the proton drip line nucleus (i.e., the lightest proton-bound nucleus) in each isotopic chain. The arrow and hollow circle for $^{17}$Na are derived from the mass values given in Ref.~\cite{Brown:2017} and Ref.~\cite{Fortune:2018}, respectively, whereas the result of the present work is denoted by a solid circle.}
\label{fig:MED_3p}
\end{figure}

The difference discussed between the observed and predicted energies of the $^{17}$Na g.s.\ is not unique.~Similar lowering of the g.s.\ energy was observed in other 3\textit{p}-emitters, such as $^{31}$K and $^{20}$Al. Such a difference may be partly explained by the effect of Thomas-Ehrmann shift (TES)~\cite{Thomas:1952,Ehrman:1951} which is often associated with single-particle quantum states of low orbital angular momentum in \textit{p}-unbound nuclei. Given that the lowering of the g.s.\ of above-mentioned 3\textit{p} emitters is remarkable, this may indicate some additional general structure feature common for these \textit{p}-emitting nuclei. 

Using the $S_p$ of $^{17}$Na and other 3\textit{p} emitters, we examined the MED systematics to probe the mirror symmetry~\cite{Fortune:2018}. Figure~\ref{fig:MED_3p} presents the MED versus the difference between neutron number $N$ and proton number $Z$ for the B, Na, Al, and K isotopes. The MED trend shows a flat pattern with decreasing $N\text{--}Z$ until the proton drip line.~Beyond the proton drip line, the MED starts to decrease, exhibiting a prominent drop in systematics for all studied 3\textit{p} emitters ($^{31}$K, $^{20}$Al, $^{17}$Na, and $^7$B), although the magnitude of the drop in $^7$B is small.~Such a regular lowering points out one common source: isospin symmetry violation.~The mechanism of mirror-symmetry breaking is often attributed to the TES effect in \textit{p}-unbound nuclei.~A recent study using the GSM concluded that the TES originates mainly from reduced Coulomb energies in the \textit{p}-unbound nuclei due to the extended radial density distributions of valence protons~\cite{Xing:2025}. Indeed, all studied 3\textit{p} emitters exhibit an extended proton density distribution compared to the neutron density distribution of their mirror partners according to the DRHBc calculations that reasonably reproduce the lowering of MEDs observed at 3\textit{p} emitters (see Appendix C for details).~The decrease in MEDs for the studied 3\textit{p} emitters (except $^7$B) is dramatic. Other 3\textit{p} emitters like $^{13}$F, $^{27}$Cl, or $^{35}$Sc may add valuable information to the MED systematics, which calls for the dedicated measurements.

In summary, the g.s.\ of 3\textit{p}-unbound isotope $^{17}$Na has been proposed.~Based on a detailed analysis of the angular correlations of all decay products, we suggest a 3\textit{p}-decay energy of 2.24($^{+0.17}_{-0.25}$)~MeV for the $^{17}$Na g.s., which is significantly smaller than the previously reported upper limit value of 4.85~MeV. The mass excess of $^{17}$Na, derived from the measured $^{17}$Na g.s.\ decay energy, is 32.11(25)~MeV. Two different theoretical interpretations have been invoked to explain the observed 2.24-MeV peak in the $^{17}$Na spectrum.~The DRHBc calculations where the $s$-wave component of the valence proton dominates describe the g.s.\ energy well.~According to the GCC calculations, the observed $^{17}$Na g.s.\ peak might be attributed to a virtual-like threshold resonance in the $^{16}$Ne\text{--}\textit{p} system governed by the 2$s_{1/2}$ pole. Such an ambiguity may be resolved by descriptions of the shape of the $^{16}$Ne\text{--}\textit{p} spectrum from the $^{17}$Na g.s.\ decay, which calls for future measurements with large statistics. Most of the theoretical models predict that the $^{17}$Na g.s.\ has a spin-parity of 1/2$^+$, suggesting a possible isospin symmetry breaking in the mirror pair $^{17}$Na\text{--}$^{17}$C. The observed effect of lowering of the $^{17}$Na g.s.\ is similar to those detected in other 3\textit{p}-emitters like $^{31}$K and $^{20}$Al, which points to a general effect in nuclear structure beyond the proton drip line.~The systematics of the MED values between known 3\textit{p} emitters and their \textit{n}-rich mirror nuclei show a regular dramatic decrease compared to the \textit{p}-bound nuclei. Such a phenomenon of ``stabilization" observed at 3\textit{p} emitters may be attributed to the continuum effects and extended proton density distributions.

\textit{Acknowledgments}\----This Letter was partially supported by the Helmholtz International Center for FAIR (HIC for FAIR); the Chinese Academy of Sciences President’s International Fellowship Initiative (Grant No. 2024PVA0005\_Y1); the National Key Research and Development Program (MOST 2022YFA1602303 and MOST 2023YFA1606404), the National Natural Science Foundation of China (Grants No. 12347106, No. 12147101, and No. 12305125); the National Key Laboratory of Neutron Science and Technology (Grant No. NST202401016); the Helmholtz Association (Grant IK-RU-002); the Russian Science Foundation (Grant No. 26-12-00228); the Polish National Science Center (Contract No. 2019/33/B/ST2/02908); the Helmholtz-CAS Joint Research Group (Grant HCJRG-108); the Ministry of Education \& Science, Spain (Contract No.\ PID2024-159209NB-C21); the Ministry of Economy, Spain (Grant FPA2015-69640-C2-2-P); the Hessian Ministry for Science and Art (HMWK) through the LOEWE funding scheme; the Justus-Liebig-Universit\"at Giessen (JLU) and the GSI under the JLU-GSI strategic Helmholtz partnership agreement; and MICIU PID2023-147569NB-C21. This work was carried out in the framework of the Super-FRS Experiment Collaboration in the FAIR project.

\begin{figure}[t]
\begin{center}
\includegraphics[width=0.45\textwidth]{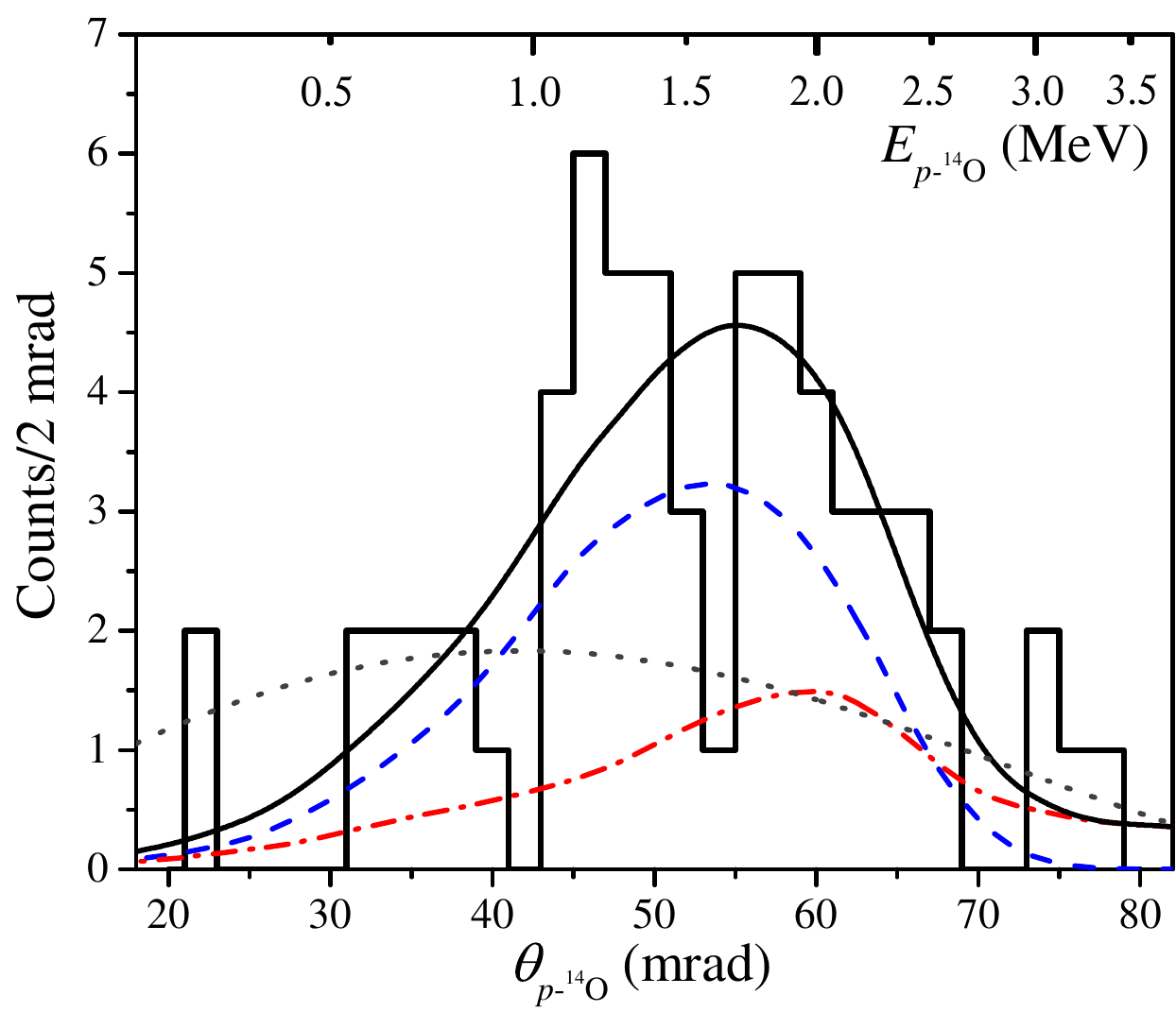}
\end{center}
\caption{Angular $\theta_{p-^{14}\rm{O}}$ correlations (histogram) derived from the measured $^{14}$O+3\textit{p} coincidences using the selection gate (iii) shown in Fig.~\ref{fig:17Na-rho3}(a) in the $84<\rho_3<100$ mrad range. The corresponding $1p$-decay energies $E_{p-^{14}\rm{O}}$ are given by the upper axis. The simulated  primary 1\textit{p} decay of the $^{17}$Na$^{*}$ state into the first-excited 2$^+$ state in $^{16}$Ne ($Q_{2p}$=3.2~MeV) with the $Q_{1p}$ of 2.04~MeV is shown by the dash-dotted curve. The contribution of secondary-emitted protons from $^{16}$Ne$^{*}$(2$^+$) into the $^{14}$O g.s.\ measured in Ref.~\cite{Mukha:2010} is shown by the dashed line. The solid curve is their sum. The dotted curve denotes the upper-limit estimate of the nonresonant contribution using four-body phase-volume simulations.}
\label{fig:theta-pNe_exc}
\end{figure}
\textit{Appendix A: Assignments of the 5.2-MeV excited state in $^{17}$Na}\----The angular $\theta_{p-^{14}\rm{O}}$ correlations obtained by imposing the $\rho_3$ gate (iii) in Fig.~\ref{fig:17Na-rho3}(a) are shown in Fig.~\ref{fig:theta-pNe_exc}. Such a selection is aimed at an excited state in $^{17}$Na located around $E_T\simeq5.2$~MeV. One may see a bump dominating the distribution at the angles of 50\text{--}60 mrad. This bump can be attributed only to the sequentially-emitted protons from $^{17}$Na$^*$ proceeding via an intermediate state in $^{16}$Ne$^{*}$ ($Q_{2p}$=3.2~MeV) and its subsequent 2\textit{p} decay to the $^{14}$O g.s.\ An energy-preferred 1\textit{p}-decay branch into the $^{16}$Ne g.s.\ and its subsequent 2\textit{p} decay should strongly contribute to the correlations at angles around 35 mrad (see the dashed curve in Fig.~\ref{fig:theta-p-Ne}), which is not supported by the data shown in Fig.~\ref{fig:theta-pNe_exc}. Such a decay selectivity may be attributed to a dominant orbital angular momentum $L_p$=0 of the first emitted proton from the 5.2-MeV excited state in $^{17}$Na into the excited state $^{16}$Ne$^{*}$(2$^+$). In such a case, another 1\textit{p}-decay branch into the $^{16}$Ne g.s.\ 0$^+$ should have $L_p$=2. Then the estimated ratio of the respective penetration factors is about 2. This value supports the $L_p$=0 decay branch into $^{16}$Ne$^{*}$(2$^+$), which is actually observed in our data. The angular correlations from the 2\textit{p}-decay branch of the first excited state 2$^+$ in $^{16}$Ne were measured and described in Ref.~\cite{Mukha:2010}. The corresponding simulation taken from Fig.~22 of Ref.~\cite{Mukha:2010} is shown in Fig.~\ref{fig:theta-pNe_exc} (dashed curve normalized to the data). The contribution of the first-emitted proton is denoted by the dash-dotted curve. The sum of these two components matches the data with maximum probability of 0.99 when the energy of the first-emitted proton is 2.04($^{+0.60}_{ -0.56}$)~MeV. This 1\textit{p}-decay energy and its uncertainties are derived similarly to those of the $^{17}$Na g.s.\ Taking the known decay energy (3.2~MeV) of the $^{16}$Ne$^{*}$(2$^+$) state into account, the derived total 3\textit{p}-decay energy is 5.24($^{+0.60}_{-0.56}$)~MeV. The described 1\textit{p}-decay with ${L_p}$=0 into the 2$^+$ state in $^{16}$Ne allows for a tentative assignment of spin-parity of the 5.2-MeV state in $^{17}$Na to be (3/2$^+, 5/2^+$).


\textit{Appendix B: GSM and GCC calculations for $^{17}$Na}\----Since $^{17}$Na is an unbound 3$p$ emitter, it is essential to account for continuum effects explicitly.~The Gamow shell model provides a state-of-the-art framework for describing open quantum systems~\cite{Michel2021book,Michel:2002}. It employs the Berggren basis~\cite{Berggren:1968} which incorporates continuum effect. Here, we use many-body perturbation theory~\cite{Coraggio2020} to derive valence-space effective Hamiltonians and transition operators in the Berggren basis~\cite{Hu2020,Xu2023,Zhang2023,Xu2025}. The g.s.\ of $^{14}$O serves as GSM cores and the valence space is $sd$-shell. (The complex-momentum contours for $\pi s_{1/2}$, $\pi d_{5/2}$, and $\pi d_{3/2}$ partial waves are defined as $k=0 \rightarrow 0.4-0.2i \rightarrow 0.8 \rightarrow 4~\text{fm}^{-1}$, $k=0 \rightarrow 0.5-0.1i \rightarrow 1.0 \rightarrow 4~\text{fm}^{-1}$, and $k=0 \rightarrow 0.65-0.20i \rightarrow 1.3 \rightarrow 4~\text{fm}^{-1}$, respectively. Each partial wave contour is discretized with 35 scattering states.) The lowest states in $^{15}$F and $^{16}$Ne are calculated first, showing good agreement with available experimental data. Then the states in $^{17}$Na are predicted and the results are summarized in Table~\ref{tab:spectrum}.
\vspace{-2mm}
\begin{table}[!htbp]
\caption{GSM theoretical predictions for resonance states in $^{17}$Na. Decay energies $E$ are given relative to the $^{14}$O g.s.\ and the units are in MeV. Widths $\Gamma$ are in keV.}
\begin{ruledtabular}
\begin{tabular}{llll}
\multirow{1}{*}{Nucleus} & \multirow{1}{*}{$J^\pi$} & \multirow{1}{*}{$E$ (MeV)} & \multirow{1}{*}{$\Gamma$ (keV)} \\
\hline
\multirow{3}{*}{$^{17}$Na} 
& $5/2^+$ & 4.252 & 837 \\
& $1/2^+$ & 4.508 & 734 \\
& $3/2^+$ & 4.739 & $\sim$0 \\
\end{tabular}
\end{ruledtabular}
\label{tab:spectrum}
\end{table}

\begin{figure}[t]
\begin{center}
\includegraphics[width=0.45\textwidth]{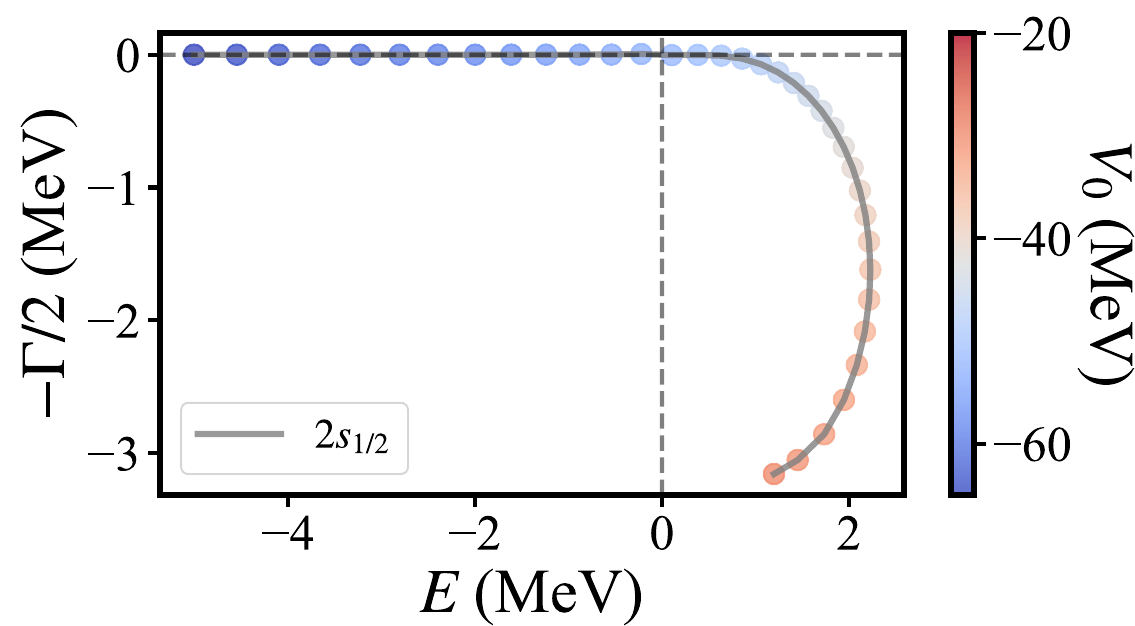}
\end{center}
\caption{GCC-calculated trajectory of the 2$s_{1/2}$ pole as a function of the Woods-Saxon potential depth $V_0$, from bound state to virtual-like threshold resonance in a $^{16}$Ne+\textit{p} configuration.~The color scale on the right-hand side is related to $V_0$.}
\label{GCC_Na17}
\end{figure}

The present GSM predictions cannot reproduce the observed near-threshold peak located around \(1~\text{MeV}\) above the \(^{16}\mathrm{Ne}+p\) threshold.~Owing to its large width and the deviation from the evaluated isobaric masses, the $^{17}$Na g.s.\ structure may reflect some anomalous feature. To clarify its origin, we employ the Gamow coupled-channel approach~\cite{Wang:2017,Wang2019}.~In the calculations, \(^{16}\mathrm{Ne}\) is an inert core with a fully occupied \(1s_{1/2}\) resonant orbital, consistent with earlier experimental evidence~\cite{Mukha:2010}. The complex-momentum contour for the \(\pi s_{1/2}\) partial wave is taken as $k = 0 \rightarrow 0.1 - 0.3i \rightarrow 0.3 - 0.3i \rightarrow 0.6 - 0.4i \rightarrow 0.8 \rightarrow 1.2 \rightarrow 4~\mathrm{fm}^{-1}$.~Each segment was discretized with 30 points.~Since the \(1s_{1/2}\) orbital is already occupied by the valence protons in \(^{16}\mathrm{Ne}\), we focus on the \(2s_{1/2}\) 1\textit{p}-orbital as a function of the Woods-Saxon potential depth.~The corresponding trajectory is shown in Fig.~\ref{GCC_Na17}. The \(2s_{1/2}\) state evolves from a genuine resonance into a threshold (nearly virtual) resonance~\cite{Michel2006PRC,Wang2019,Pfutzner:2023,Charity:2023,Yang2023a},~appearing at $Q_p\lessapprox2$~MeV. Such (virtual-like) threshold resonances generally exhibit widths comparable to or exceeding their decay energies but located very close to or even below the threshold. However, this relation can be modified by configuration mixing and the spectroscopic factor.

\textit{Appendix C: DRHBc calculations for 3\textit{p} emitters and their mirror partners}\----To investigate the possible reason of the lowered MEDs of 3\textit{p} emitters, the g.s.\ properties of $^{17}$Na and other 3\textit{p} emitters ($^7$B, $^{20}$Al, and $^{31}$K) as well as their mirror nuclei are calculated using the DRHBc theory.~While this theory assumes the box boundary condition for nuclear systems, the continuum effects can be taken into account by expanding wave functions in a Dirac Woods-Saxon basis~\cite{Zhou2003PRC}, which has an appropriate asymptotic behavior for describing the possible large spatial extension in exotic nuclei. Consequently, the DRHBc theory has been widely applied to a number of nuclear-structure phenomena~\cite{Zhang2022ADNDT,Guo2024ADNDT}, and successfully to explore proton halo nuclei~\cite{Zhang2024PRC,Zhang2025PLB} and proton emitters~\cite{Lu2024PLB,Zhang2025PRC(L)}. The DRHBc-predicted separation energies of 3\textit{p} emitters and their mirror partners together with corresponding experimental values are presented in Table~\ref{tab:DRHBc_sep}. It is evident that the data are reasonably reproduced by the calculations. Especially, the predicted $S_p$ value for $^{17}$Na lies very close to the data.~Meanwhile, the DRHBc results can provide a quantitative interpretation of the drop in MEDs observed at 3\textit{p} emitters. For instance, the calculated MED of $2.32$~MeV for $^{17}$Na--$^{17}$C mirror pair approaches the experimental value shown in Fig.~\ref{fig:MED_3p}. 
\vspace{-3mm}
\begin{table}[!hbtp] 
\centering
\caption{Experimental and DRHBc-predicted separation energies (in MeV) for 3\textit{p} emitters and their mirror partners.}
\begin{ruledtabular}
\begin{tabular}{llllll}
\multirow{2}{*}{\textit{p}-rich} & \multicolumn{2}{c}{$S_p$ (MeV)} & \multirow{2}{*}{\textit{n}-rich} & \multicolumn{2}{c}{$S_n$ (MeV)} \\
\cmidrule(lr){2-3} \cmidrule(lr){5-6}
 & EXP & DRHBc & & EXP & DRHBc \\
\midrule
$^{7}$B & -2.013(26) & -2.54 & $^{7}$He & -0.410(8) & -1.16 \\
\midrule
$^{17}$Na & -0.84(25) & -1.20 & $^{17}$C & 0.734(18) & 1.12 \\
\midrule
$^{20}$Al & -1.17(10) & -2.43 & $^{20}$N & 2.16(8) & 1.62 \\
\midrule
$^{31}$K & -2.15(15) & -2.01 & $^{31}$Mg & 2.312(3) & 4.88 \\
\end{tabular}
\end{ruledtabular}
\label{tab:DRHBc_sep}
\end{table}

\begin{figure}[ht]
\begin{center}
\includegraphics[width=0.45\textwidth]{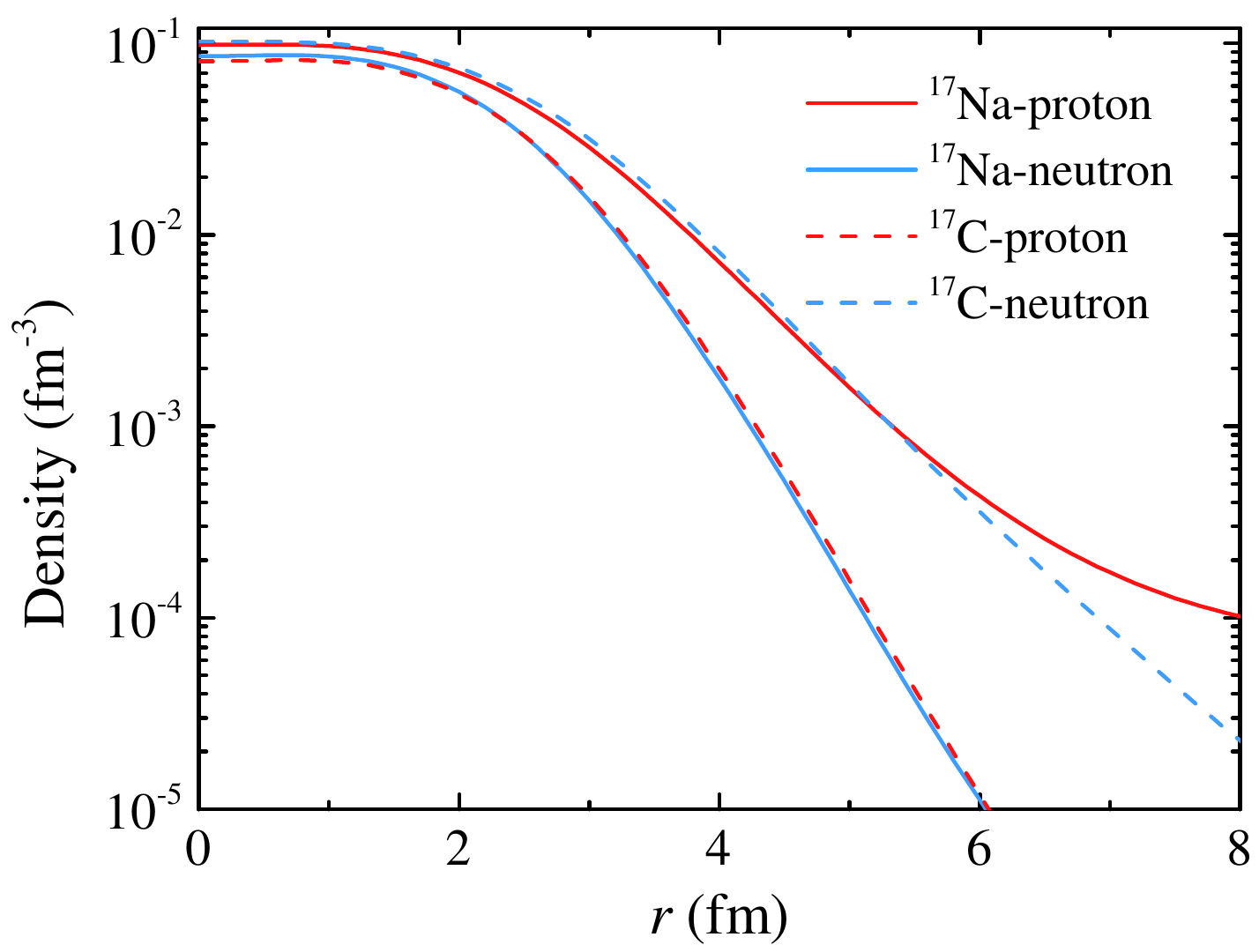}
\end{center}
\caption{Proton- and neutron-density distributions calculated with the DRHBc theory for mirror pair $^{17}$Na\text{--}$^{17}$C.}
\label{fig:dens_3p_mirror}
\end{figure}

Furthermore, we have calculated proton- and neutron-density distribution for all above-mentioned 3\textit{p} emitters and their mirror nuclei. As an example, the calculated \textit{p}- and \textit{n}-density profiles of the $^{17}$Na\text{--}$^{17}$C mirror pair are shown in Fig.~\ref{fig:dens_3p_mirror}. It is worth mentioning that the derived proton- and matter-distribution radii of $^{17}$C, based on the density profiles shown in Fig.~\ref{fig:dens_3p_mirror}, are 2.45 fm and 2.81 fm, respectively, in reasonable agreement with experimental values~\cite{Kanungo2016PRL,Dobrovolsky2021NPA}. Due to the dominant $s$-wave component of the valence proton, the proton density distribution of $^{17}$Na is more extended to large distances compared to the neutron density distribution of $^{17}$C. In contrast, $^{17}$Na's \textit{n}-density distribution is nearly identical to $^{17}$C's \textit{p}-density distribution. Similar behavior is observed for all other studied 3\textit{p} emitters. It was proposed that the Coulomb energy difference between the mirror pairs can largely account for the MED, see Eq. (5) in Ref.~\cite{Yu2024PRL}. In the present DRHBc calculations for the $^{17}$Na\text{--}$^{17}$C mirror pair, the MED extracted from the Coulomb energies, which are calculated with the proton density distributions shown in Fig.~\ref{fig:dens_3p_mirror}, is 2.58~MeV. This value can explain the lowering of experimental MED shown in Fig.~\ref{fig:MED_3p} to a large extent. Therefore, an extended \textit{p}-density distribution of proton-unbound nuclei results in a decrease of Coulomb repulsion, thereby lowering the g.s.\ energy.
\\

\bibliographystyle{apsrev4-2}
\bibliography{arXiv_ref}
\end{document}